\documentclass[letterpaper,english,aps,pre,reprint]{revtex4-1}
\usepackage[T1]{fontenc}
\usepackage[latin9]{inputenc}
\usepackage{color}
\usepackage{sidecap}
\usepackage{amsmath}
\usepackage{amssymb}
\usepackage{graphicx}
\sidecaptionvpos{figure}{c}
\makeatletter

\pdfpageheight\paperheight
\pdfpagewidth\paperwidth

\providecommand{\tabularnewline}{\\}

\makeatother

\usepackage{babel}
\begin{document}

\title{Structure, size and statistical properties of chaotic components
in a mixed-type Hamiltonian system}

\author{\v{C}rt Lozej}
\email{clozej@gmail.com}

\affiliation{CAMTP - Center for Applied Mathematics and Theoretical Physics, University
of Maribor, Mladinska 3, Maribor, Slovenia}

\author{Marko Robnik}
\email{marko.robnik@guest.um.si}

\affiliation{CAMTP - Center for Applied Mathematics and Theoretical Physics, University
of Maribor, Mladinska 3, Maribor, Slovenia}

\date{\today}
\begin{abstract}
We preform a detailed study of the chaotic component in mixed-type
Hamiltonian systems on the example of a family of billiards (introduced
by Robnik 1983). The phase space is divided into a grid of cells and
a chaotic orbit is iterated a large number of times. The structure
of the chaotic component is discerned from the cells visited by the
chaotic orbit. The fractal dimension of the border of the chaotic
component for various values of the billiard shape parameter is determined
with the box counting method. The cell filling dynamics is compared
to a model of uncorrelated motion the so-called random model (Robnik
\emph{et. al.} 1997) and deviations attributed to sticky objects in
the phase space are found. The statistics of the number of orbit visits
to the cells is analyzed and found to be in agreement with the random
model in the long run. The stickiness of the various structures in
the phase space is quantified in terms of the cell recurrence times.
The recurrence time distributions in a few selected cells as well
as the mean and standard deviation of recurrence times for all cells
are analyzed. The standard deviation of cell recurrence time is found
to be a good quantifier of stickiness on a global scale. Three methods
for determining the measure of the chaotic component are compared
and the measure is calculated for various values of the billiard shape
parameter. Lastly, the decay of correlations and the diffusion of
momenta is analyzed. 
\end{abstract}
\maketitle

\section{\label{sec:introduction}Introduction}

Generic Hamiltonian systems are typically neither integrable nor fully
chaotic, but exhibit both regular and chaotic motion depending on
the initial condition. The phase space is divided into several invariant
components where the dynamics is regular on some and chaotic on others.
The structure of the chaotic component in such mixed-type systems
may be very complex, with the chaotic sea enveloping multitudes of
KAM islands. The islands influence the dynamics of the chaotic orbits
in their vicinity. Chaotic orbits that come close to a KAM island
may spend extended periods of time in its vicinity in quasi regular
motion, a phenomenon known as stickiness (Refs. {[}\onlinecite{Bun2012}{]}
and {[}\onlinecite{Cont2010}{]} provide a good introduction to the
topic). Additionally, cantori \citep{Mackay1984a,Mackay1984}, i.e.
the fractal remains of destroyed invariant tori, may also be present
in the chaotic sea, acting as barriers to the chaotic dynamics. These
types of intermittent behavior in Hamiltonian systems with mixed phase
space may lead to the dynamics being only weakly chaotic in the sense
that time correlations as well as recurrence time distributions decay
more slowly \citep{Chirikov1984,Chirikov1999,VCG1983,Zaslavsky2002,Cristadoro2008,Abud2013},
non-exponentially as opposed to the exponential decay in some chaotic
systems with strong mixing properties \citep{Young1998,Young1999,Hirata1999}.
It is however important to stress that a divided phase space is not
a prerequisite for slow decay of correlation. An example of an ergodic
K-system with power law decay of correlations is the stadium billiard
\citep{Bun1979,markarian2004}. There the slow decay of correlations
has been linked to the presence of sticky bouncing ball and boundary
glancing orbits \citep{VCG1983}.

Both the complexity of the phase space and the intermittent behavior
of the orbits pose a challenge to determining the size or relative
measure of a chaotic component. The typical island around island structures
lead to a fractal border of the chaotic component. Determining the
measure of the chaotic component to a high degree of accuracy may
thus require very detailed phase portraits. These are usually obtained
by iterating a chaotic orbit and letting it explore all of the available
phase space. The problem is that the orbits may become frequently
trapped by sticky objects and a large number of iterations may be
needed to fully resolve the phase portrait. Cantori may also limit
the accessibility to some parts of the chaotic component.

The need to accurately determine the measure of the chaotic component
was motivated by the study of energy level statistics in the context
of quantum chaos in generic autonomous Hamiltonian systems. In the
strict semiclassical limit it is possible to separate chaotic eigenstates
from regular ones (see Refs. \citep{Percival1973,Berry1977,voros1979,Shnirelman1979,BerryRobnik1984,Rob1998,veble1999,BR2010,BR2013,BR2013A,BLR2017}
and references therein). The full spectrum can then be decomposed
as a linear superposition of the chaotic and regular spectrum. The
relative size of each component is given by the relative measure of
the component in the classical phase space. An accurate value for
the size of the chaotic component in the equivalent classical system
is therefore vital in analyzing the spectral statistics of mixed-type
quantum systems in the semiclassical limit. 

The main goal of this paper is to present a method of determining
the measure of the largest chaotic component in a mixed-type system,
using as an example the family of billiards introduced in Ref. {[}\onlinecite{Rob1983}{]}.
The method involves dividing the phase space into a grid of cells
and counting the visits of a chaotic orbit in each of the cells. The
statistics of the orbit visits is then analyzed and compared to a
model of completely uncorrelated random motion. If the cells dividing
the phase space are sufficiently small and the number of orbit iterations
suficiently large to reach the ergodic limit, the measure of the chaotic
component can be extracted with a high degree of accuracy.

The paper is organized as follows. In Sec. \ref{sec:Structure} we
define the family of billiards studied and describe some of the features
of their phase spaces. We also describe the division of the phase
space into cells. In Sec. \ref{sec:Fractal} we estimate the fractal
dimension for various billiards belonging to the aforementioned family.
In Sec. \ref{sec: Filling} we analyze the filling of the cells with
regard to the number of orbit iterations as well as the statistics
of the number of cell visitations and compare them with a model of
completely uncorrelated motion. In Sec.\ref{sec:RetTimes} we analyze
the cell recurrence times and their statistics. In Sec. \ref{sec:Measure}
we compare three methods of determining the measure of the chaotic
component and discuss the connection to quantum chaos. In Sec. \ref{sec:Correlation}
we examine the momentum autocorrelation function and the momentum
diffusion. In Sec. \ref{sec:Discussion} the results are discussed
and concluded.

\section{\label{sec:Structure}The structure of the chaotic components}

\begin{figure}[h]
\begin{centering}
\includegraphics{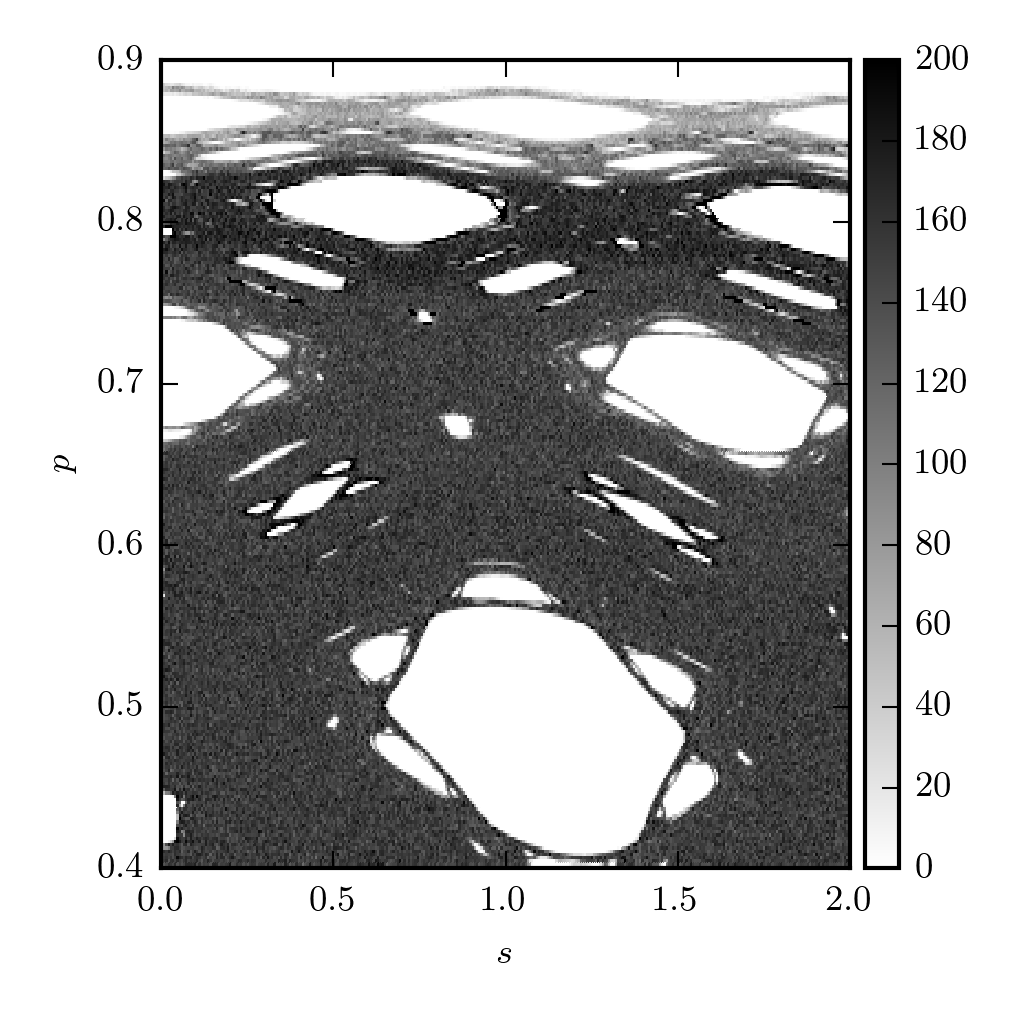}
\par\end{centering}
\caption{\label{fig:Portrait}Part of the phase space visited by a single chaotic
orbit in the billiard defined by Eq. \eqref{eq:Billiard} for $\lambda=0.135$.
The number of visits is represented by the lightness of the plot.}
\end{figure}

As an example of a mixed-type system we chose to study the family
of billiards introduced in Ref. {[}\onlinecite{Rob1983}{]}. The border
of the billiard can be described in the complex plane as a smooth
conformal mapping of the unit circle $|z|=1,$
\begin{equation}
z\rightarrow z+\lambda z^{2},\label{eq:Billiard}
\end{equation}
where the shape parameter is a real number $\lambda\in\left[0,\,0.5\right]$.
The billiard border is a circle at $\lambda=0$ and thus integrable.
It has been proven that in the other extreme case $\lambda=0.5$ the
billiard is an ergodic K-system \citep{Mar1993}. Between both extremes
the phase space is mixed containing both chaotic and regular components.
The shape of the border is convex up to $\lambda=0.25$ and concave
for larger values of the shape parameter. The phase space is defined
by the Poincar\'e-Birkhoff coordinates $\left(s,\,p\right)$, where
$s$ is the arc-length of the billiard boundary in the counterclockwise
direction measured from the point where it intersects the real axis,
and the momentum is the sine of the reflection angle of the particle
$p=\sin\alpha$. The phase space is thus a cylinder $x=\left(s,\,p\right)\in\left[0,\,\mathcal{L}\right]\times\left[-1,\,1\right]$,
where we take $s$ to be periodic with a period equal to the total
length of the billiard boundary 
\begin{equation}
\mathcal{L}=\int_{0}^{2\pi}\mathrm{d}\varphi\sqrt{1+4\lambda^{2}+4\lambda\cos\varphi}.
\end{equation}
The billiard map, mapping the successive collisions, is area preserving
\citep{Ber1981}. In Fig. \ref{fig:Portrait} we show a chaotic orbit
in a small region of the phase space for the $\lambda=0.135$ billiard.
The shade of gray denotes the number of times the orbit has visited.
Unvisited areas are depicted in white. Because of the multitude of
KAM islands of stability, the chaotic component has a complex fractal
structure as is typical for mixed-type Hamiltonian systems. The borders
of the islands can also represent sticky objects. The chaotic orbit
tends to stick to the area around an island for an extended number
of bounces. We may also see larger areas of a mostly uniform shade
distinct from those of other areas, most notably on the upper border
of the chaotic component. This indicates the presence of cantori,
the remnants of the destroyed invariant KAM tori, that act as barriers
for the orbit. Another notable feature of strictly convex billiards
as is the case for $\lambda<0.25$ is the presence of Lazutkin caustics
\citep{Laz1973}. A curve is called a caustic of a billiard when,
if some link of a trajectory (by link we mean the straight segment
of the trajectory between two consecutive bounces) is tangent to the
curve, then all other links of the same trajectory are tangent to
this curve. The Lazutkin caustic is manifested in the phase space
as a Lazutkin torus. The Lazutkin tori limit the area accessible to
the chaotic component near the upper and lower border of the phase
space. In Fig. \ref{fig:Portrait} the torus is reached at $p\approx0.9.$
We note that at the shape parameter value $\lambda=0.25$ the Lazutkin
tori are destroyed \citep{mather1982} because a point with zero curvature
exists on the billiard border at $s=\frac{\mathcal{L}}{2}$, $z=-1$. 

\begin{figure}[h]
\begin{centering}
\includegraphics[width=1\columnwidth]{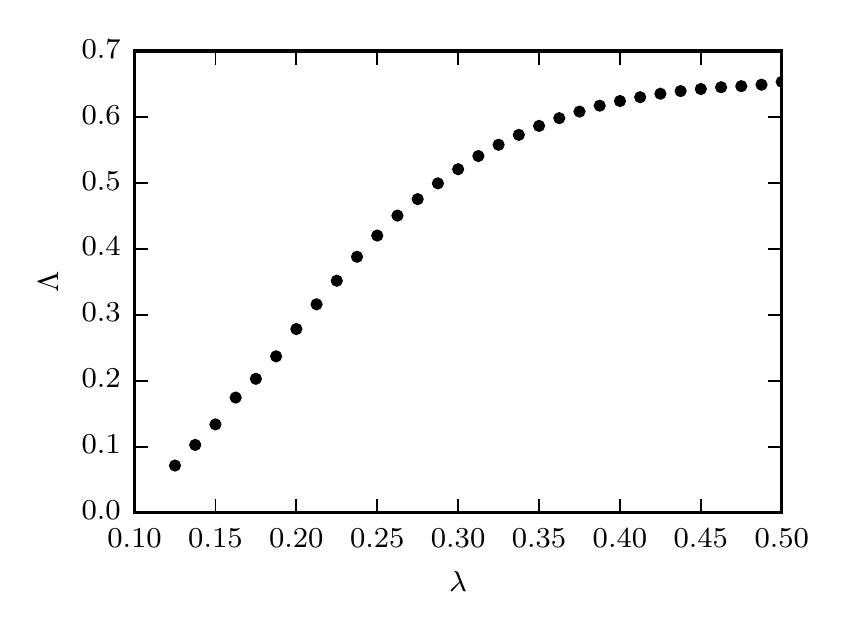}
\par\end{centering}
\caption{L\label{fig:Lyapunov}yapunov exponent on the largest chaotic component
as a function of the shape parameter.}
\end{figure}

We will concentrate our analysis on the billiards with a single dominant
chaotic component much larger in size than the other chaotic components.
In terms of the shape parameter this means $\lambda\geq0.1175$. For
lower values multiple small chaotic components form around the unstable
periodic orbits and are separated by invariant tori. These are comparable
in size and each represents only a small fraction of the entire phase
space. At $\lambda\approx0.1175$ most of the tori are destroyed and
several chaotic components merge into a single larger one. To confirm
that the orbits in these areas of phase space are indeed chaotic we
determined the Lyapunov exponent. Let $\delta x\left(T\right)$ be
the separation of two orbits at $T$ iterations and $\delta x_{0}$
the initial separation. The Lyapunov exponent is then
\begin{equation}
\varLambda=\lim_{T\rightarrow\infty}\lim_{\delta x_{0}\rightarrow0}\frac{1}{T}\ln\frac{\left|\delta x\left(T\right)\right|}{\left|\delta x_{0}\right|}.
\end{equation}
 The Lyapunov exponent on the largest chaotic component as a function
of the shape parameter, determined by the method introduced in Ref.
\citep{Benettin1976}, is presented in Fig. \ref{fig:Lyapunov}. 

\begin{figure}[h]
\begin{centering}
\includegraphics[width=1\columnwidth]{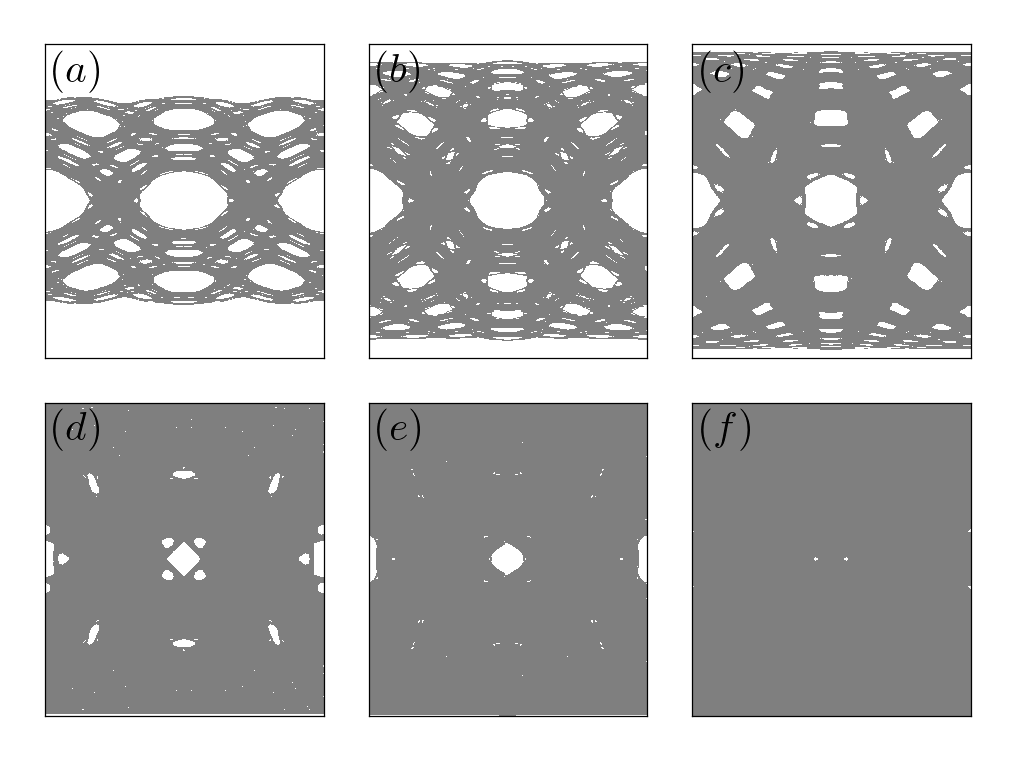}
\par\end{centering}
\caption{\label{fig:ChaoticComponent}Portraits of the largest chaotic component
in the phase space cylinder$\left(s,\,p\right)$ for different values
of the shape parameter $\lambda$. (a) $\lambda=0.1175$, (b) $\lambda=0.135$, (c) $\lambda=0.15$, (d) $\lambda=0.18$, (e) $\lambda=0.2$, (f) $\lambda=0.25$. The gray area represents the chaotic
component.}
\end{figure}

In Fig. \ref{fig:ChaoticComponent} we show the area of phase space
occupied by the largest chaotic component for various values of the
shape parameter. At $\lambda=0.135$ the accessible phase space is
still significantly limited by the Lazutkin tori in addition to the
large number of stable islands. When $\lambda$ is increased the tori
are moved ever further towards the border of the phase space, until
finally disappearing at $\lambda=0.25$. Furthermore, the billiards
become increasingly chaotic with more and more of the stable islands
being destroyed until only some tiny islands remain at $\lambda=0.25$.
At $\lambda=0.5$ the billiard is rigorously proven to be ergodic
\citep{Mar1993}. The question arises: At a given shape what proportion
of the phase space is filled by the chaotic component? The complexity
of the structure of the phase space as well as the presence of sticky
objects pose significant obstacles in answering this question. Because
the border of the chaotic component is fractal it is difficult to
decide whether an area on the border of the phase space still belongs
to the chaotic component or not. Chaotic orbits may also become trapped
by sticky objects and cantori and may need a very large number of
iterations to explore all of the available phase space.

To determine the sizes of the chaotic components we improved the procedure
from Ref. {[}\onlinecite{Dobnikar1996}{]}. The phase space is partitioned
on a rectangular grid into $L\times L$ cells. We then select an initial
condition in the largest chaotic component of the billiard and iterate
the orbit for a large number of bounces $T$. We monitor which of
the cells are visited by the chaotic orbit to determine whether the
portion of the phase space contained in the cell is a part of the
chaotic component or not. We will refer to cells that are never visited
by the chaotic orbit as \emph{empty cells} and cells that are visited
as \emph{filled cells}. 

\section{\label{sec:Fractal}Fractal dimension of the boundary}

To measure the complexity of the boundary of the chaotic component
we calculated its fractal dimension \citep{Ott2002} using the box-counting
method. To determine the fractal dimension of a set, we cover the
set with boxes of some side length $\epsilon$. Let $N_{B}\left(\epsilon\right)$
be the number of boxes needed to cover the set at a given $\epsilon$.
The box-counting dimension is 
\begin{equation}
D=\underset{\epsilon\rightarrow0}{\lim}\frac{\log N_{B}\left(\epsilon\right)}{\log\left(\frac{1}{\epsilon}\right)}.\label{eq:BoxDim}
\end{equation}
In the case of our billiards we use the phase space cells as defined
above for the boxes. The side length of a cell is proportional to
the inverse of the grid size $\epsilon\propto\frac{1}{L}$. We refer
to cells that are filled but have at least one empty neighboring cell
as \emph{border cells}. The orbit visited a border cell at least once,
meaning at least some of the phase space contained in it is part of
the chaotic component. However, it is likely that the whole cell is
not filled by the chaotic component because the neighboring cells
are empty. The border of the chaotic component is thus covered by
the border cells. From the definition \eqref{eq:BoxDim} we derive
the following relation for the number of border cells $\log N_{B}=D\log L+\mathrm{const}.$
We can determine the fractal dimension by counting the number of border
cells at various grid sizes and fitting the data with a line.

\begin{figure}[h]
\begin{centering}
\includegraphics[width=1\columnwidth]{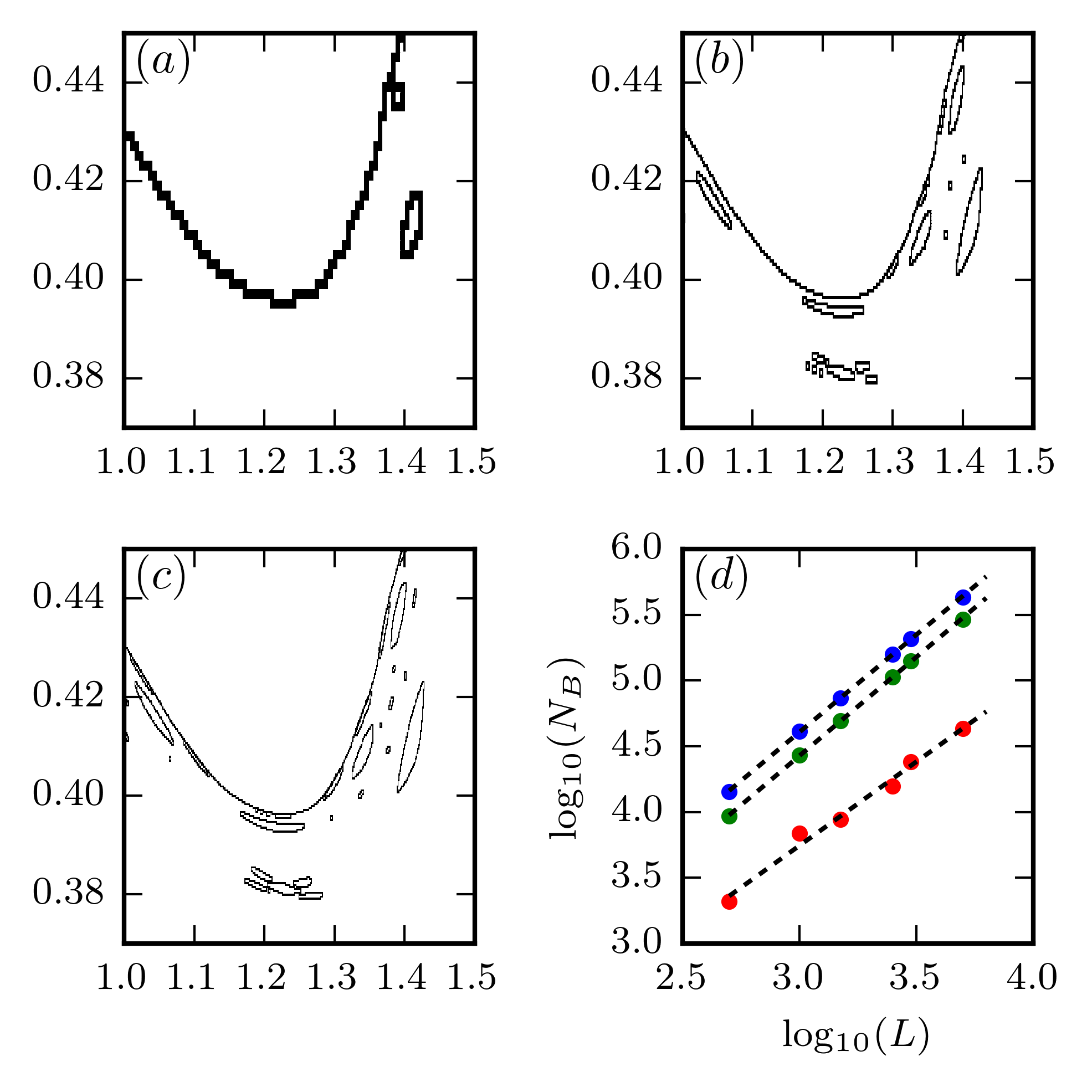}
\par\end{centering}
\caption{\label{fig:FractalDim}Plots showing the border cells of the largest
chaotic component in the vicinity of a stable island in the $\lambda=0.15$
billiard at various grid sizes (a) $L=1000$, (b) $L=3000$ and (c)
$L=5000$. Panel (d) shows the logarithm of the number of border cells
versus the logarithm of the grid size for three different values of
$\lambda$. From top to bottom (blue) $\lambda=0.1175$, (green) $\lambda=0.15$
and (red) $\lambda=0.2$. The slope of the line is the estimate for
the fractal dimension.}
\end{figure}

In determining the fractal dimension we used cell grids between $L=500$
and $L=5000$. The number of orbit iterations was $T=10^{11}$ ensuring
a large number of iterations per cell even for the largest grid size.
In Fig. \ref{fig:FractalDim} (a-c) we show the border cells for a
small area of phase space near one of the islands around the stable
period three orbit of the $\lambda=0.15$ billiard. As the cells get
smaller more and more details of the structure of the border of the
chaotic component may be discerned. The logarithm of the number of
border cells versus the logarithm of the grid size for three different
values of $\lambda$ is shown on panel (d) of the same figure. We
fit the data with a least squares linear regression model. The slope
of the line is the fractal dimension $D$. The values of the fractal
dimension along with an error estimate for various $\lambda$ are
given in Table \ref{tab:FractalDimensions}. In the cases of larger
values $\lambda>0.2$ this can be considered as only a rough estimate,
because of the very small number of tiny stable islands. The overall
number of border cells is therefore small and significant fluctuations
occur when varying the grid size. To get a more accurate estimate,
larger grid sizes would be needed, but this is prohibitive due to
the need to further increase the number of orbit iterations. For all
$\lambda$ values the fractal dimension has a value significantly
above $1$, indicating that the structure of the border is indeed
very complex. Because of this complexity, border cells may pose a
significant contribution to the overall area of the chaotic component.

\begin{table}[h]
\begin{centering}
\begin{tabular}{|c|c||c|c|}
\hline 
$\lambda$ & $D$ & $\lambda$ & $D$\tabularnewline
\hline 
\hline 
0.1175 & $1.48\pm0.01$ & 0.16 & $1.48\pm0.03$\tabularnewline
\hline 
0.125 & $1.52\pm0.01$ & 0.18 & $1.62\pm0.05$\tabularnewline
\hline 
0.13 & $1.48\pm0.01$ & 0.20 & $1.27\pm0.08$\tabularnewline
\hline 
0.135 & $1.42\pm0.01$ & 0.23 & $1.68\pm0.47$\tabularnewline
\hline 
0.14 & $1.46\pm0.01$ & 0.25 & $1.77\pm0.60$\tabularnewline
\hline 
0.15 & $1.50\pm0.01$ & 0.5 & 1\tabularnewline
\hline 
\end{tabular}
\par\end{centering}
\caption{\label{tab:FractalDimensions}Table of the fractal dimensions for
border of the chaotic component. The error is estimated from the error
of the linear regression slope. The value 1 indicates that all of
the phase space is filled by the chaotic component.}

\end{table}

\section{\label{sec: Filling}The filling of cells and the occupancy distribution}

The simplest estimate for the measure of the chaotic component is
counting the number of filled cells. This gives us an upper estimate,
provided the \emph{number of iterations is large enough} for the orbit
to visit all of the cells that are at least partially filled by the
chaotic component. In a previous paper \citep{LozRob2018} we have
shown that the \emph{random model} introduced in Ref. {[}\onlinecite{Rob1997}{]}
describes the filling of the cells well for ergodic systems exemplified
by the stadium billiard \citep{Bun1979}. One would expect that the
motion of the orbit on the chaotic component of a mixed type system
should be ergodic on sufficiently long time scales and the random
model should be applicable. 

The random model assumes an uncorrelated Poissonian filling of the
cells. Let $\chi_{A}$ be the proportion of cells that will be filled
in the infinite time limit. Ideally, for sufficiently small cells
this would be equal to the measure of the chaotic component. We will
denote the measure of the chaotic component by $\chi_{C}$. The random
model predicts that the proportion of filled cells as a function of
the number of orbit iterations has the following form
\begin{equation}
\chi\left(T\right)=\chi_{A}\left(1-\exp\left(-\frac{T}{N_{C}}\right)\right),\label{eq:CellFill}
\end{equation}
where $N_{C}=\chi_{A}L^{2}$ is the number of cells available to the
chaotic orbit. Let us next define the cell occupancy $M$ as the number
of times a cell was visited by the orbit. Following the previous assumption
that the orbit visits are completely random, the probability that
a cell is visited in the next iteration is equal for all cells and
given by the normalized cell size $a=1/N_{C}$. The probability that
a cell will be visited $M$ times in $T$ iterations is given by the
binomial distribution \citep{Feller1968}
\begin{equation}
P_{B}\left(M\right)=\begin{pmatrix}T\\
M
\end{pmatrix}a^{M}\left(1-a\right)^{T-M}.
\end{equation}
The mean and variance of the binomial distribution are
\begin{equation}
\begin{array}{cc}
\mu=\left\langle M\right\rangle =Ta, & \sigma^{2}=\left\langle \left(M-\mu\right)^{2}\right\rangle =\mu(1-a),\end{array}
\end{equation}
respectively. In the limit of small cells $a\rightarrow0$ and large
numbers of iterations $T\rightarrow\infty$, but with $\mu=Ta=\mathrm{const.}$
the binomial distribution converges toward the Poissonian distribution
\begin{equation}
P_{P}\left(M\right)=\frac{\mu^{M}\mathrm{e^{-\mu}}}{M!}.\label{eq:Poisson}
\end{equation}
The variance of the Poissonian distribution is equal to its mean $\sigma^{2}=\mu$.
The cell occupancy distribution should converge to the Poissonian
distribution with $\mu=\frac{T}{\chi_{C}L^{2}}$ in the long time
and small cell limit.

\begin{figure}[h]
\begin{centering}
\includegraphics[width=1\columnwidth]{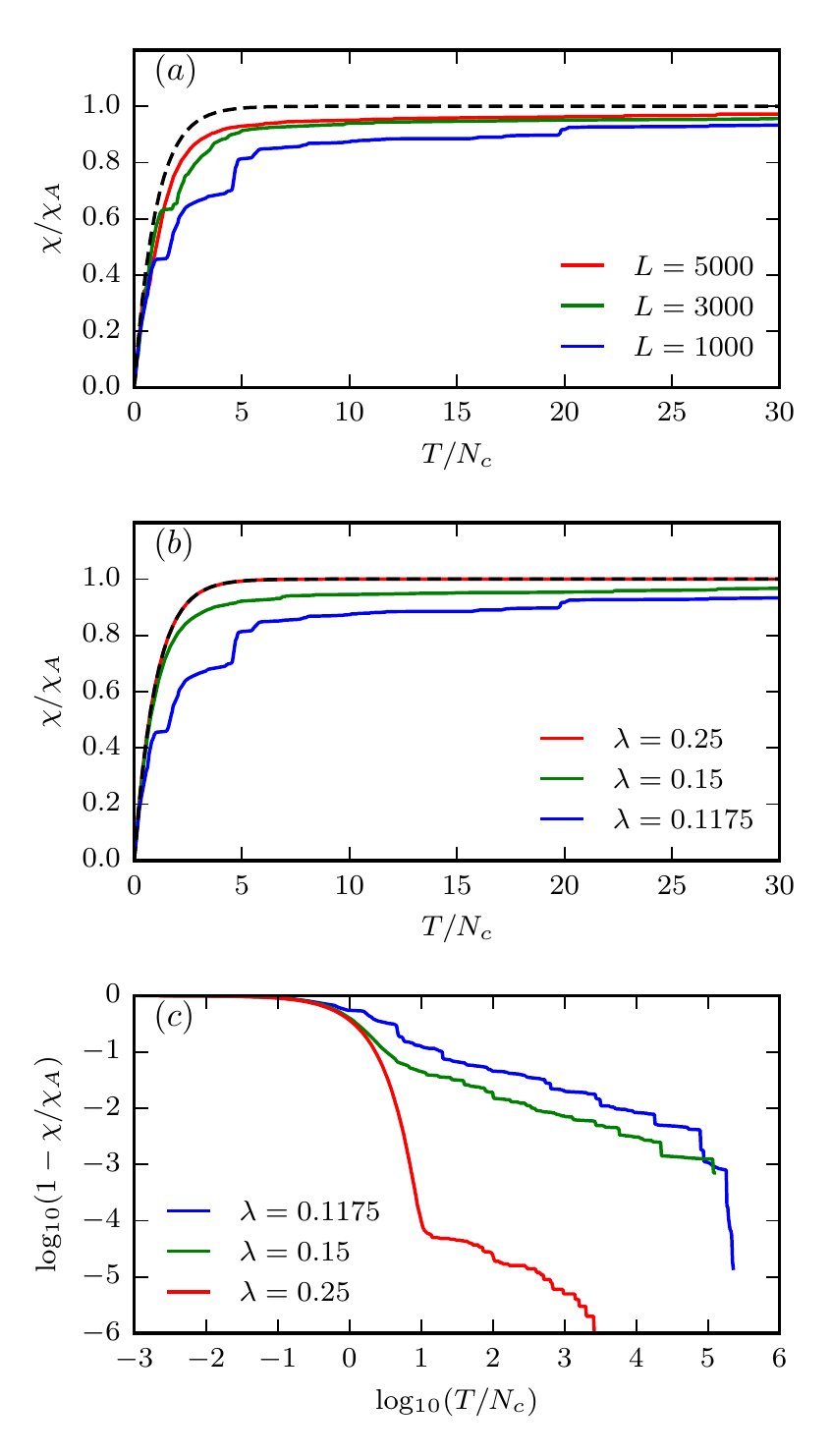}
\par\end{centering}
\caption{\label{fig:CellFill}Filling of cells as a function of the number
of orbit iterations (a) for different grid sizes at $\lambda=0.1175$
and (b) for different values of $\lambda$ at grid size $L=1000$.
The proportion of filled cells is normalized to the asymptotic value
at $T=10^{11}$ iterations and the iterations are given in terms of
the number of chaotic cells $N_{C}=\chi_{A}L^{2}$. The black dashed
curve shows the prediction of the random model Eq. \ref{eq:CellFill}.
(c) The difference of the proportion of filled cells to the asymptotic value in the
log-log scale for the same values of $\lambda$ and grid size as in panel (b) (the order of the curves from top to bottom is indicated by the legend in each panel).}

\end{figure}

In Fig. \ref{fig:CellFill} we show how the proportion of filled cells
normalized to the asymptotic value changes with the number of iterations.
The number of iterations is given in units of $N_{C}$. Panel (a)
shows the cell filling for $\lambda=0.1175$ for three grid sizes.
The cell filling is slower than the random model prediction Eq. \eqref{eq:CellFill}
for all three cases with visible plateaus in the curves. The plateaus
are most evident in the curve for the smallest grid size $L=1000$.
In panel (b) the cell filling is shown for three values of $\lambda$
with $L=1000$. At $\lambda=0.1175$ and $\lambda=0.15$ the cell
filling is still slower than expected from the random model whereas
for $\lambda=0.25$ the cell filling follows the prediction quite
closely. The plateaus in this case $\lambda=0.25$ become visible
only when the difference of the proportion of filled cells and the
asymptotic value is of the order of less than $10^{-4}$. This may
be seen in panel (c) of Fig. \ref{fig:CellFill} where the logarithmic
plots of the cell filling for the same three values of $\lambda$
are shown at grid size $L=1000$.

It is clear that the assumption that the filling of the cells is completely
uncorrelated does not hold for the chaotic component of a mixed type
system. The plateaus in the cell filling curve may be explained with
the presence of sticky objects and cantori. Because the orbit may
become trapped in a region of phase space by a cantorus or in the
vicinity of a sticky object for an extended number of iterations,
some cells are visited more often than expected. The number of filled
cells stops increasing for the duration of the trapping. When the
orbit eventually escapes the sticky object, the number of filled cells
starts increasing until the orbit is captured again. The exact structure
of the plateaus is different for orbits with different initial conditions
but the general shape of the curve is the same. Long plateaus with
small rises suggest that some small areas of the phase space are particularly
inaccessible. One such area is the border of the chaotic component
near the Lazutkin caustics that contains both many KAM islands as
well as cantori.

Smaller grid sizes $L$ (larger cells) cause the cell visits to be
more correlated. If the cells are large enough for a sticky object
to be covered by only a few cells, these cells are visited a large
number of times in rapid succession. When the cell size is decreased
(larger $L$), more cells are needed to cover the sticky object and
the visits are more evenly distributed. Billiards with larger values
of $\lambda$ also exhibit less correlated cell visits as the number
of KAM islands of significant size, that are the main source of stickiness,
is greatly reduced. At $\lambda=0.5$ the billiard is ergodic and
exhibits no stickiness. The cell filling curve has no plateaus and
follows the random model prediction precisely. 

\begin{figure}[h]
\begin{centering}
\includegraphics[width=1\columnwidth]{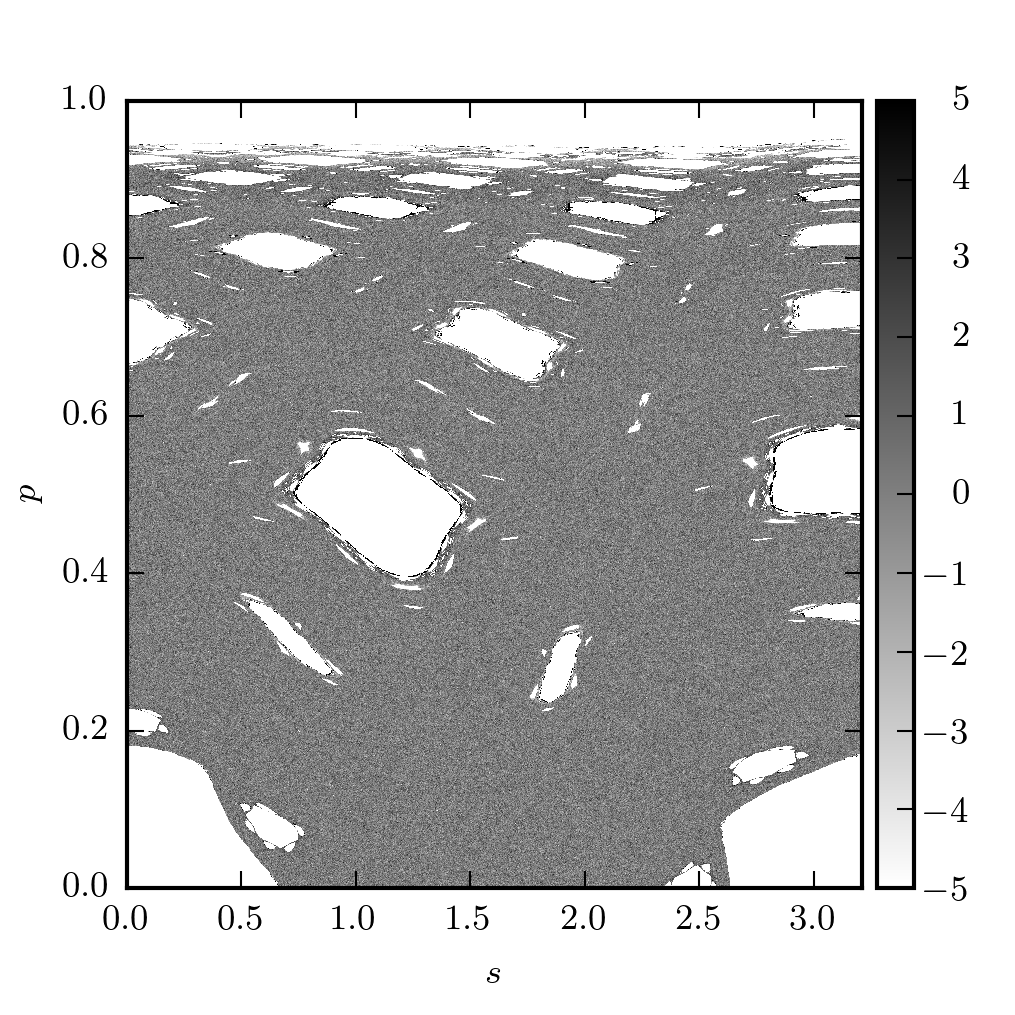}
\par\end{centering}
\caption{\label{fig:CellOccupancy}Cell occupancy in one quadrant of the phase
space for $\lambda=0.15$, $L=3000$ and $T=10^{11}$. The grayness
indicates the occupancy as the difference to the average $\mu=\frac{T}{\chi_{C}L^{2}}$.
The scale is given in units of the standard deviation $\sigma=\sqrt{\mu}$.}
\end{figure}

It is useful to consider not only whether a cell gets filled but also
its occupancy, meaning the number of times it is visited. A plot of
the cell occupancy for the $\lambda=0.15$ billiard after $10^{11}$orbit
iterations is presented in Fig. \ref{fig:CellOccupancy}. Because
of symmetry we show only one quadrant of the phase space. Inside the
chaotic sea the cell occupancy is quite uniform with the majority
of cells having an occupancy close to $M\approx\mu=\frac{T}{\chi_{C}L^{2}}$,
within the expected standard deviation given by the random model $\sigma=\sqrt{\mu}$.
Notable differences occur in the occupancy of the cells on the borders
of the KAM islands and near the Lazutkin tori. In these one can see
a difference of more than $5\sigma$ from the average occupancy. Unusually
large occupancy numbers indicate areas with sticky objects. Conversely,
areas with unusually small occupancy indicate that transport into
these areas is limited by some barrier like for instance a cantorus.
Once the orbit penetrates the area bordered by a cantorus it may be
traped there for some time. 

\begin{figure*}[t]
\begin{centering}
\includegraphics[width=1\textwidth]{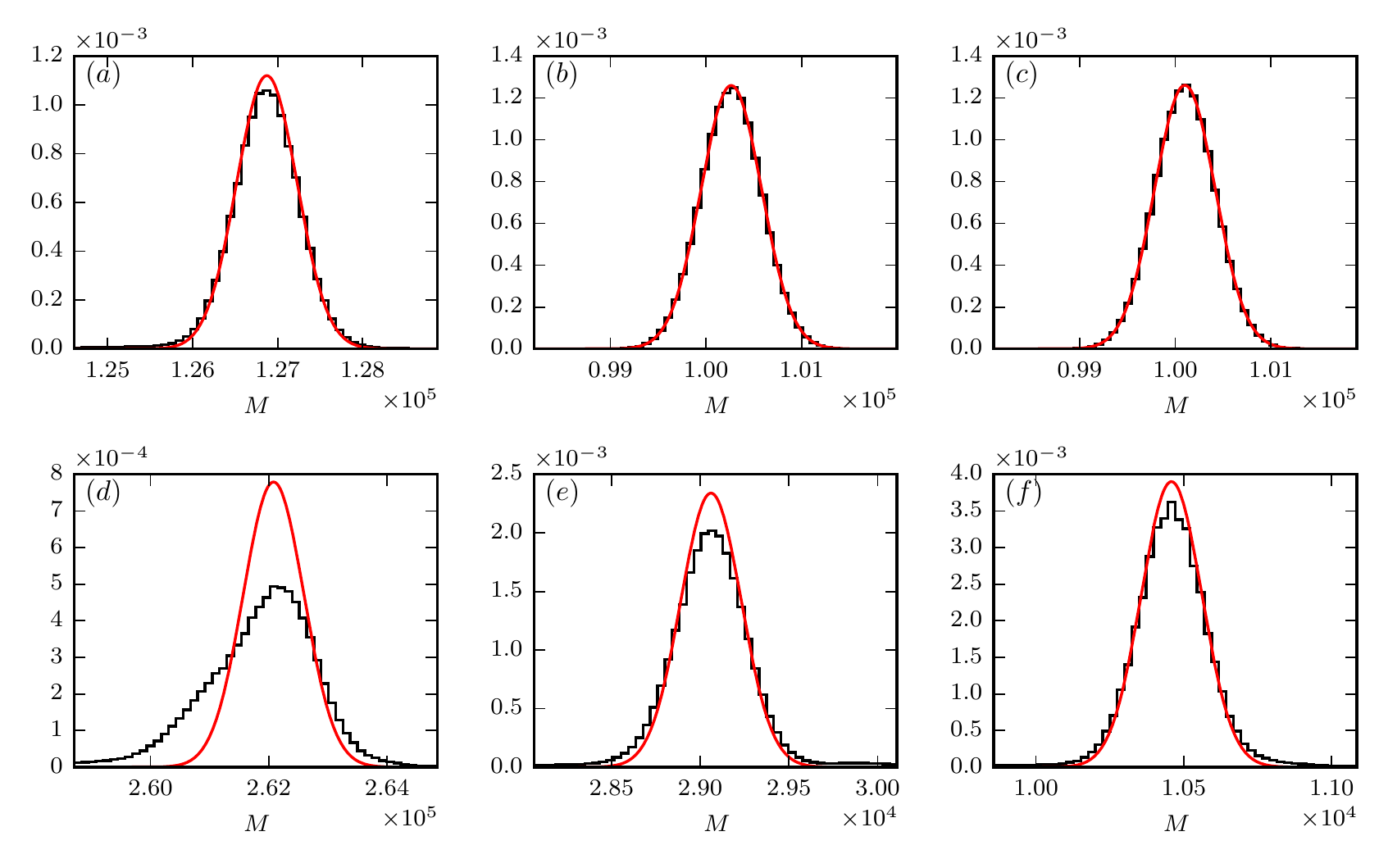}
\par\end{centering}
\caption{\label{fig:OccupancyDist}Histograms of the occupancy distributions. Top row $L=1000$ and 
(a) $\lambda=0.15$ , (b) $\lambda=0.25$ , (c) $\lambda=0.5$. Bottom row $\lambda=0.1175$ and (d) $L=1000$,
(e) $L=3000$, (f) $L=5000$. The Poissonian
distribution Eq. \eqref{eq:Poisson}, expected from the random model,
is fitted to the histograms and shown with the red curve. The number
of orbit iterations is $T=10^{11}$.}
\end{figure*}

The occupancy distributions of the cells after $10^{11}$ orbit iterations
are presented in Fig. \ref{fig:OccupancyDist}. Panels (a), (b), (c)
show the distributions for $L=1000$ and $\lambda=0.15$, $\lambda=0.25$,
$\lambda=0.5$, respectively. The histograms are fitted with the Poissonian
distribution Eq. \eqref{eq:Poisson} which closely coincides with
the histograms in these three cases. On the other hand the histogram
for $\lambda=0.1175$ and $L=1000$ in panel (d) is markedly different
from the Poissonian distribution and exhibits some asymmetry with
regard to the mean value. The asymmetry is greatly diminished if the
cell size is decreased and the histograms are again close to Poissonian
as shown in panels (e) and (f). We stress that the only difference
in panels (d), (e) and (f) is the grid of cells, the orbit is exactly
the same. The occupancy distributions also have a peak of height $1-\chi_{A}$
at $M=0$ because of the empty cells. The exception is $\lambda=0.5$
where, due to ergodicity all cells are filled.

The fact that the cell occupancy distributions are close to Poissonian
implies that the cell visits are in the long term uncorrelated, provided
the cells are small enough. Even though the chaotic orbit may spend
a significant number of iterations trapped in the vicinity of sticky
objects this is averaged out in the long term. This suggests that
the time between successive trappings is longer than the time of the
trapping episodes. The more complex the phase space the longer it
takes (in terms of the number of orbit iterations) to reach the Poissonian
limit. If the phase space is simple like for instance in the $\lambda=0.25$
and $\lambda=0.5$ the limiting occupancy distribution is reached
before $10^{10}$ orbit iterations for $L=1000$. In contrast, for
the most complex case $\lambda=0.1175$ the distribution is far from
Poissonian even after $10^{11}$ orbit iterations. Decreasing the
cell size (increasing $L$) generally improves the rate of convergence.
The average occupancy of the cells may be used to estimate the measure
of the chaotic component, since the relation $\mu=\frac{T}{\chi_{C}L^{2}}$
should hold. The naive estimate from the number of filled cells $\chi_{A}$
and from the average of the occupancy distribution $\chi_{C}$ are
slightly different, as will be shown in more detail in Sec. \ref{sec:Measure}.

\section{\label{sec:RetTimes}Cell recurrence times}

Further information about the structure of the phase space may be
gained from the (discrete) cell recurrence times. Recurrence times
statistics are one of the standard ways of quantifying stickiness
\citep{Altman2004,Altman2005,Altman2006,Abud2013}. We define the
cell recurrence time $\tau$ for each individual cell as the number
of iterations an orbit needs to return to the same cell for the first
time. If the motion on the chaotic component is ergodic, it follows
from the Kac lemma \citep{Kac1959}, that the average first return
time $\mu_{\tau}$ to a cell is equal to the inverse of the normalized
measure of the cell, which is equal the number of cells in the chaotic
component $\mu_{\tau}=\chi_{C}L^{2}=N_{C}$. Additionally, if the
cell visits are completely uncorrelated (random model) the probability
that a cell is visited after any number of iterations is equal. The
probability distribution of cell return times is thus exponential
\begin{equation}
P\left(\tau\right)=\frac{1}{\mu_{\tau}}\exp\left(-\frac{\tau}{\mu_{\tau}}\right),\label{eq:ExponentialDist}
\end{equation}
The variance of this distribution is $\sigma_{\tau}^{2}=\mu_{\tau}^{2}$. 

Now we will take into account that the samples in our numerical experiments
are finite. We define the normal distribution as 
\begin{equation}
\mathcal{N}\left(x,\mu,\sigma\right)=\frac{1}{\sqrt{2\pi\sigma^{2}}}\exp\left(-\frac{\left(x-\mu\right)^{2}}{2\sigma^{2}}\right),
\end{equation}
 where $\mu$ and $\sigma$ are the mean and standard deviation of
the normal distribution. Let $x$ be some observable distributed according
to the probability distribution $P\left(x\right)$ with a finite mean
$\mu_{x}$ and variance $\sigma_{x}^{2}$. Let $\overline{x}$ be
the mean of $x$ as calculated from a sample of $N$ values. From
the central limit theorem \citep{Feller1968} it follows that $\overline{x}$
is distributed according to the normal distribution $\mathcal{N}\left(\overline{x},\mu_{x},\frac{\sigma_{x}}{\sqrt{N}}\right)$.
If the cell visits are uncorrelated, the distribution of the average
cell return times $\overline{\tau}$ is normal
\begin{equation}
P\left(\overline{\tau}\right)=\mathcal{N}\left(\overline{\tau},N_{C},\frac{N_{C}^{\frac{3}{2}}}{\sqrt{T}}\right),\label{eq:NormalDist}
\end{equation}
where the mean and standard deviation are determined using the exponential
distribution Eq. \eqref{eq:ExponentialDist} and the estimate that
each cell is visited $N=\frac{T}{N_{C}}$ times for the sample size.

\begin{figure}[h]
\begin{centering}
\includegraphics[width=1\columnwidth]{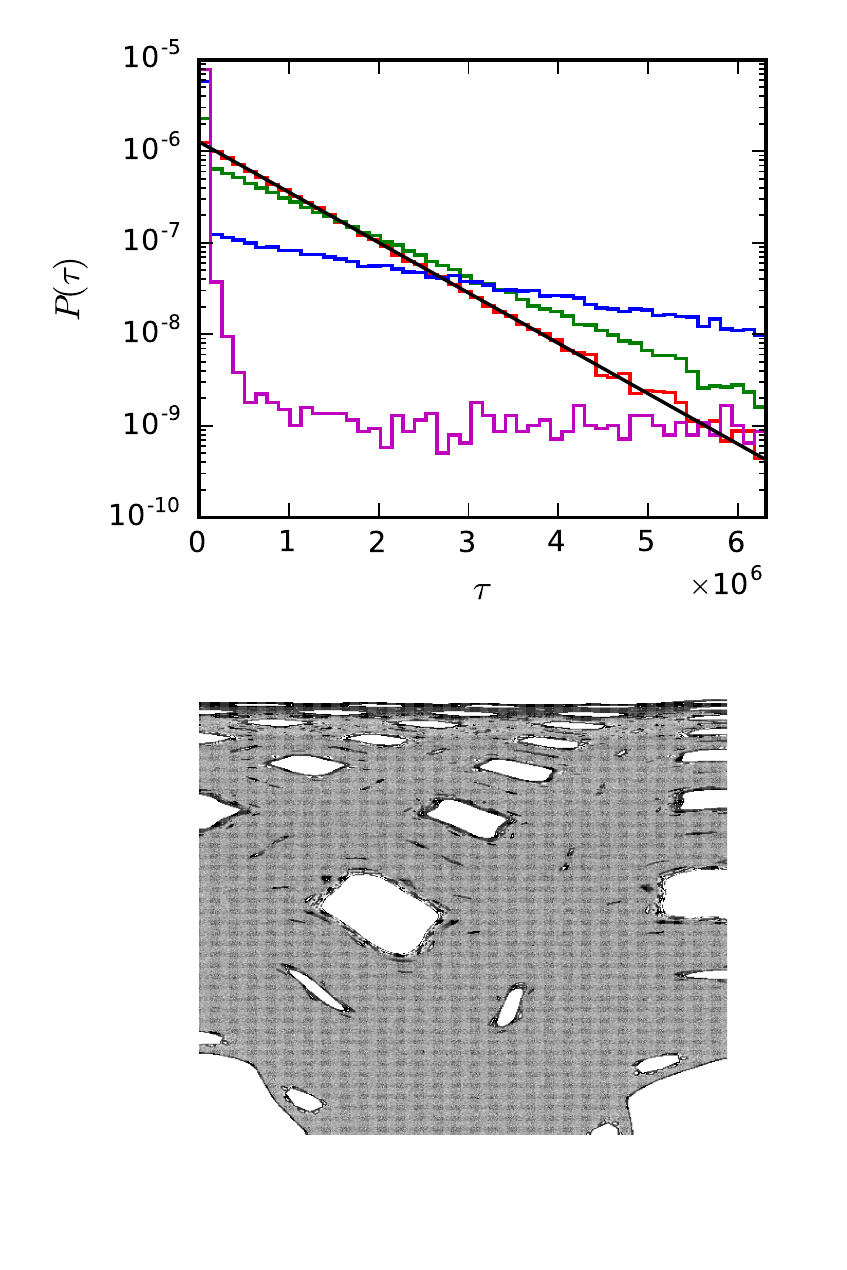}
\par\end{centering}
\caption{\label{fig:CellReturnTimes}Cell return times in the $\lambda=0.15$
billiard, with $L=1000$ and $T=10^{11}$. (a) Histograms of cell
return times for four different cells at the following coordinates
$(s,p)=(1.80,\,0.022)$ red 1, (3.14, 0.17) green 2, (1.67, 0.65) blue 3,
(2.97, 0.94) magenta 4. The exponential distribution from the random
model Eq. \ref{eq:ExponentialDist} is plotted with the black line.
(b) The average return times $\overline{\tau}$ in one quadrant of
the phase space. The shade indicates the difference from the average
$\mu_{\tau}=N_{C}$ in units of the expected standard deviation $\sigma_{\overline{\tau}}=N_{C}^{3/2}/\sqrt{T}$.
The colored and numbered boxes show the positions of the cells from panel (a) (The
size of the boxes does not correspond to the size of the cells).}
\end{figure}

The histograms of the cell recurrence times in four different cells
in the $\lambda=0.15$ billiard are presented in Fig. \ref{fig:CellReturnTimes}
(a). The first cell (red) is located in the middle of the chaotic
sea far from any KAM islands. The distribution of return times closely
coincides with the exponential distribution Eq. \eqref{eq:ExponentialDist}.
All cells sufficiently deep inside chaotic sea exhibit this kind of
exponential distribution of cell return times. This means that the
majority of cells experiences completely uncorrelated orbit visits.
The second cell (green) is located on the border of one of the largest
KAM islands. The distribution is still close to exponential with a
slightly different exponent but has a small peak at short return times.
The change in the exponential is a consequence of the increased probability
of short recurrence times due to memory effects
i. e. stickiness (see Ref. {[}\onlinecite{Altman2004}{]}). We would
like to emphasize that the recurrences to the cell are generated by
a single chaotic orbit. Even if the cell partly intersects a regular
region the orbit only visits the chaotic part of the cell. The distributions
of return times from a regular component are expected to exhibit power
law tails \citep{Buric2003,Hu2004}. The third cell (blue) is located
on the border of the smaller KAM island. The cell return time distribution
is qualitatively similar to that in the second cell but with a much
higher peak at short times and the long time tail is stronger. The
fourth cell (magenta) is located near the Lazutkin torus. The cell
return time distribution is even more severely peaked at the short
return times, while the tail is very long but still exponential. The
orbit visits this cell either in quick succession or after very long
periods of time resulting in a distribution with a large variance.
The variance (or standard deviation) of the return times in a cell
may thus be a good measure of the stickiness of a cell.

\begin{figure*}[t]
\begin{centering}
\includegraphics[width=1\textwidth]{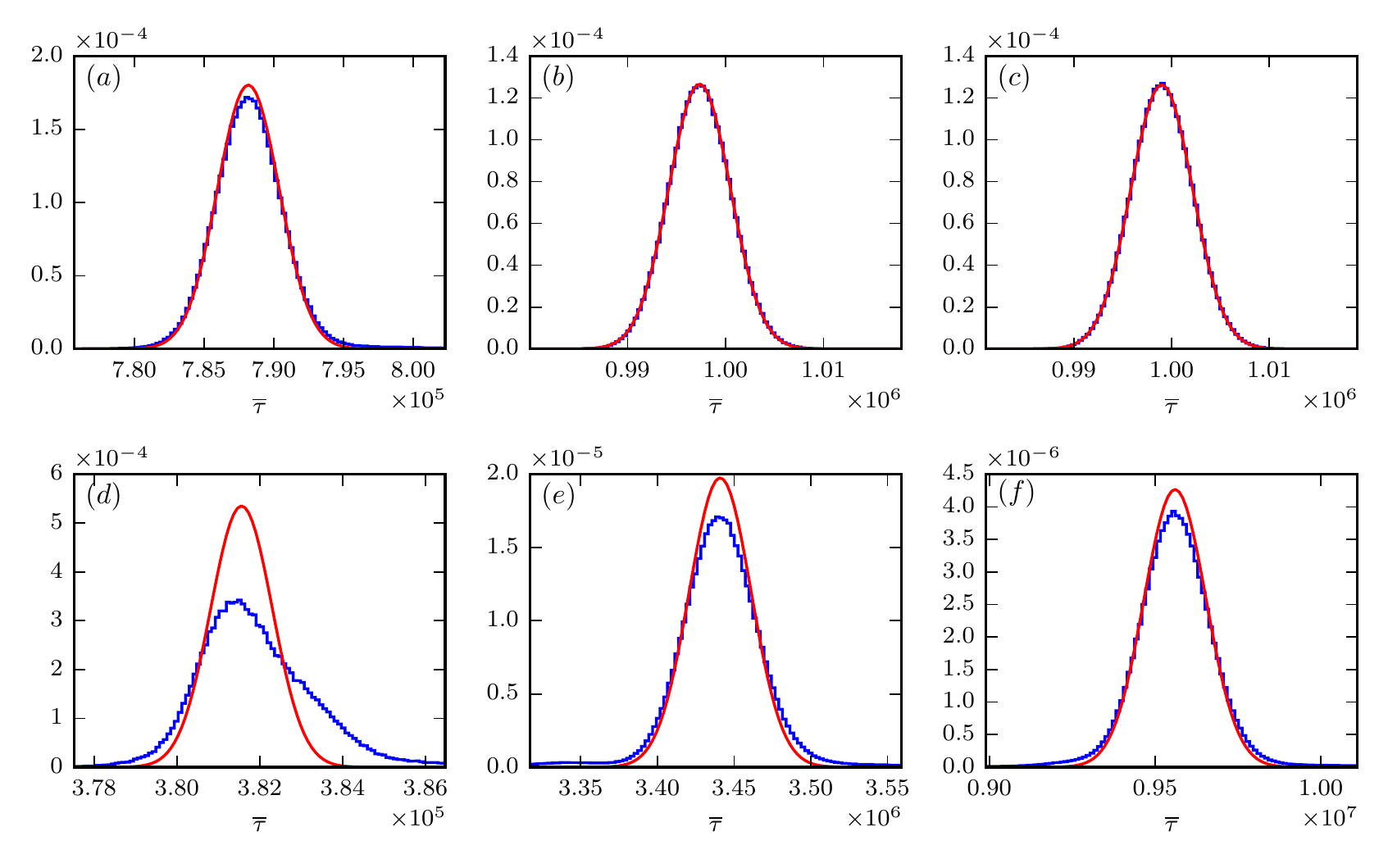}
\par\end{centering}
\caption{\label{fig:AverageReturnTimes}Histograms of the average cell return
times. Top row $L=1000$ and 
(a) $\lambda=0.15$ , (b) $\lambda=0.25$ , (c) $\lambda=0.5$. Bottom row $\lambda=0.1175$ and (d) $L=1000$,
(e) $L=3000$, (f) $L=5000$. The normal
distribution Eq. \eqref{eq:NormalDist} is plotted with the red curve.
The number of orbit iterations is $T=10^{11}$.}
\end{figure*}

To gain a global understanding of the cell return times, the average
return times in one quadrant of the phase space are shown in panel
(b) of the same figure together with the positions of the aforementioned
cells. The average cell return time is mostly uniform in the chaotic
sea. Differences of more than five standard deviations can be seen
mostly around the KAM islands and near the Lazutkin torus. The histograms
of the average return times for different sets of parameters are shown
in Fig. \ref{fig:AverageReturnTimes}. They are compared with the
random model prediction Eq. \eqref{eq:NormalDist}. We see that the
histograms agree well with the normal distribution for $\lambda=0.15$,
$\lambda=0.25$ and $\lambda=0.5$ at grid size $L=1000$, while for
$\lambda=0.1175$ there is a deviation similar to the one found in
the occupancy distribution. This deviation diminishes when the cell
size is decreased. The average cell return times hold essentially
the same information about the dynamics as the cell occupancy numbers.

\begin{SCfigure*}
\begin{centering}
\includegraphics[width=1.33\columnwidth]{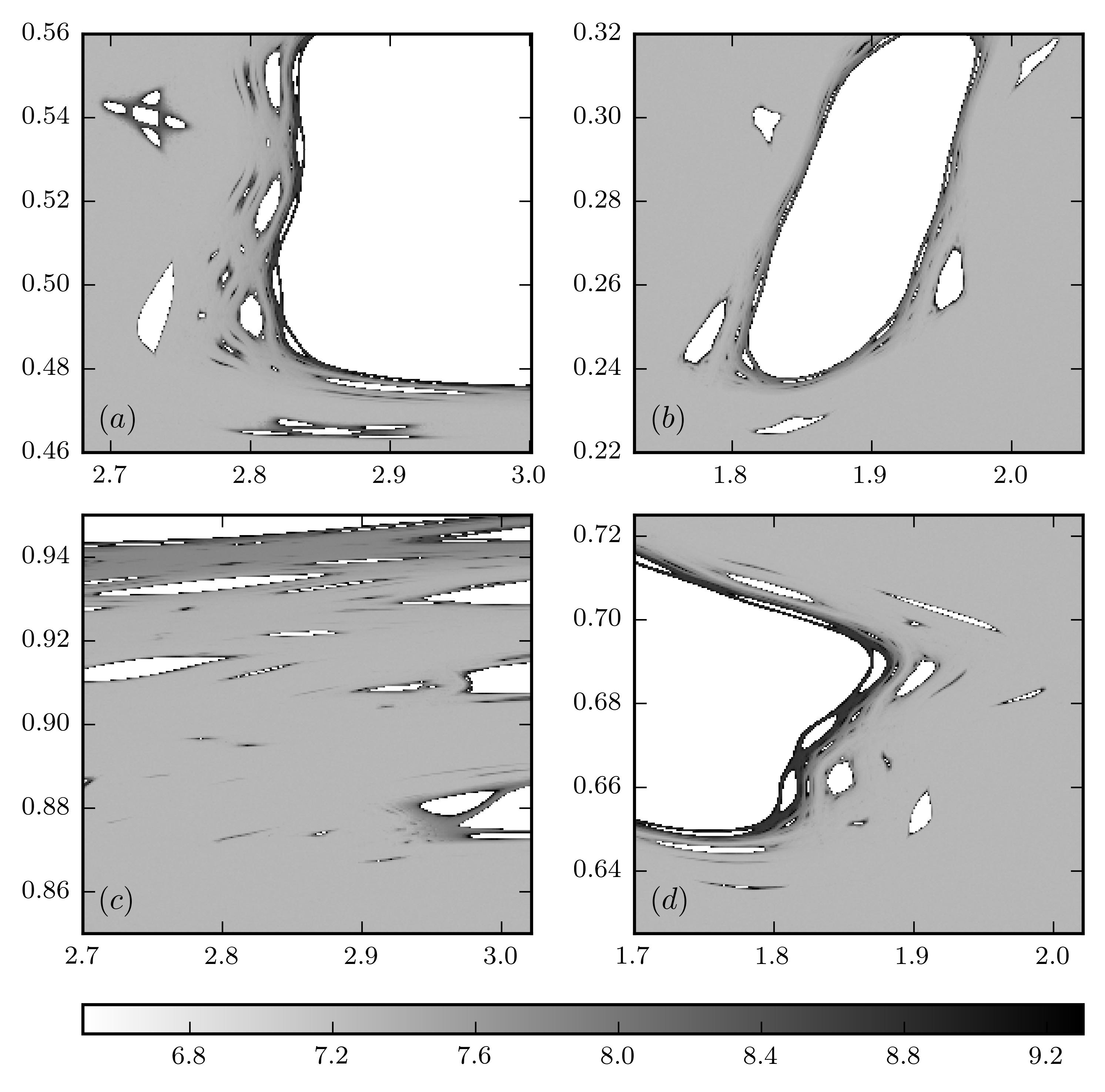}
\par\end{centering}
\caption{\label{fig:RetTimeVar}The standard deviation of cell return times
in selected areas of phase space for the $\lambda=0.15$ billiard,
with $L=5000$ and $T=10^{11}$. The greyness scale is logarithmic with
base 10.}

\end{SCfigure*}

A global quantitative assessment of the stickiness of the various
structures present in the phase space can be made from the standard
deviations (or variances) of the cell return times. Fig. \ref{fig:RetTimeVar}
shows logarithmic color plots of the cell return time standard deviations
in the $\lambda=0.15$ billiard. Panel (a) shows the phase space area
near the border of one of the KAM islands formed around the stable
periodic orbit with period three, panel (b) near a KAM island of the
star shaped period five orbit, panel (c) near the Lazutkin torus and
(d) near a KAM island of the pentagonal shaped period five orbit.
The sizes of the areas depicted are the same for each panel. In the
chaotic sea, the cell return time standard deviation is uniform, while
the structures in the phase space exhibit various degrees of stickiness.
For most of the KAM islands the standard deviation of the cell return
times increases roughly exponentially (linearly in the logarithmic
scale of the plot) in the vicinity of the border. An abrupt rather
than gradual increase in the standard deviation implies a cantorus
acting as a barrier to the orbit. This can be clearly seen in the
upper part of panel (c). Upon closer inspection one can see that the
darkest areas around the largest KAM islands in panels (a) and (d)
also have an abrupt border followed by an exponential decay. High
resolution phase portraits of the cell return time standard deviations
in the largest chaotic component for various $\lambda$ are made available
in the supplemental materials \citep{SM}. 

\section{The measure of the chaotic component\label{sec:Measure}}

\begin{figure}[h]
\begin{centering}
\includegraphics[width=1\columnwidth]{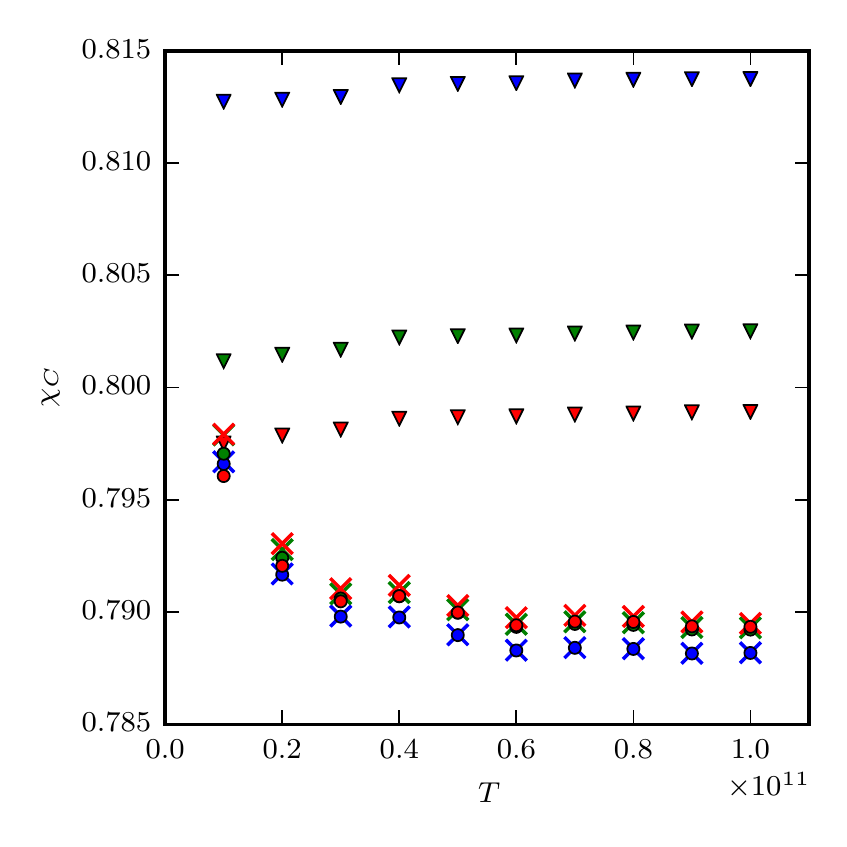}
\par\end{centering}
\caption{\label{fig:Methods}The measure of the largest chaotic component in
the $\lambda=0.15$ as determined by three different methods as a
function of the number of orbit iterations $T$ and grid size $L$.
The colors denote the grid size $L=1000$ blue, $L=3000$ green, $L=5000$
red. The markers denote the method used. Method 1 (see list in main
text) is represented by triangles, method 2 by circles and method
3 by exes. The orbit was exactly the same in all cases.}

\end{figure}

The statistical properties of the filling of the phase space cells
discussed in the previous sections may be used to determine the measure
of the chaotic components in several ways:
\begin{enumerate}
\item The simplest estimate for the measure is the proportion of filled
cells in the long time limit $\chi_{A}$. The number of cells visited
after a set number of orbit iterations is counted and divided by the
number of all cells. 
\item The measure can be obtained from the cell occupancy distributions.
Assuming the motion of the orbit on the chaotic component is ergodic
in the long term the visitation probability is distributed according
to the Poissonian distribution Eq. \eqref{eq:Poisson}, with $\mu=\frac{T}{\chi_{C}L^{2}}$.
The histogram of cell occupancies is fitted with the Poissonian and
the measure $\chi_{C}$ extracted. 
\item The measure can be obtained from the cell return times. If the motion
of the orbit is uncorrelated the probability that the mean first return
time to a cell $\overline{\tau}$ is given by the normal distribution
Eq. \eqref{eq:NormalDist}, centered at $\mu_{\tau}=\chi_{C}L^{2}=N_{C}$
and with a standard deviation of $\sigma_{\overline{\tau}}=N_{C}^{3/2}/\sqrt{T}$.
The normal distribution is fitted to the histogram of mean return
times and $\chi_{C}$ is extracted. 
\end{enumerate}
The results for the measure of the largest chaotic component in the
$\lambda=0.15$, determined by the three different methods, are compared
in Fig. \ref{fig:Methods}. The two numerical parameters in the calculation
are the number of orbit iterations $T$ and the grid size $L$. It
is advisable that the number of iterations is several decadic orders
of magnitude larger than the number of cells. Of the three methods
of determining the measure of the chaotic component method 1 has the
strongest dependence on $L$. As already mentioned some cells may
be only partially filled by the chaotic component whereas using method
1 the whole area of these cells is attributed to the chaotic component.
Because the border of the chaotic component is fractal this may lead
to a significant overestimation of its measure. Increasing $L$ and
thus improving the resolution in the phase space can significantly
decrease this overestimation. Since each cell is counted as soon as
the orbit visits it, the estimate for the measure of the chaotic component
using method 1 is an increasing function of $T$. This method may
serve as an upper estimate provided the number of orbit iterations
is large enough for the orbit to visit all the cells at least partially
containing the chaotic component. 

The benefit of method 2 is that it takes into account not only if
the cell was visited but also the number of cell visits. Cells that
are seldom visited have a high probability of being positioned near
the border and only containing a small part of the chaotic component.
The cells in the chaotic sea form the most significant contribution.
This method assumes that in the long term the motion of the orbit
on the chaotic component is ergodic and the episodes of trapping average
out. It is also important that orbit permeates all the areas of the
chaotic component bordered by strong barriers like cantori. Because
of this the number of iterations needed to achieve convergence may
be very large. In Fig. \ref{fig:Methods} we see that the estimates
for the measure at $T=10^{10}$ and $T=10^{11}$ differ significantly.
The shape of the histogram also serves as a good indication if the
grid size is sufficient. In the case of $\lambda=0.15$ the shape
is close to Poissonian already at $L=1000$ (see Fig. \ref{fig:OccupancyDist})
and increasing it to $L=3000$ has only a small effect and the results
for $L=5000$ overlap. The results of using method 3 i.e. the cell
return time histograms practically overlap with those of method 2.
This is to be expected as the mean return time to a cell is the number
of orbit iterations divided by the number of visits. 

Our main motivation for developing methods for determining the measure
of the chaotic component stems from the study of energy level statistics
in the context of quantum chaos in generic autonomous Hamiltonian
systems like for instance a quantum billiard of the same shape \citep{BR2010,BR2013,BR2013A}.
In the strict semiclassical limit it is possible to separate chaotic
eigenstates from regular ones (see Refs. \citep{Percival1973,Berry1977,voros1979,Shnirelman1979,BerryRobnik1984,Rob1998,veble1999,BR2010,BR2013,BR2013A}
and references therein). In the mixed-type regime, following the so-called
principle of uniform semiclassical condensation (of Wigner functions
of the eigenstates), Berry-Robnik (or Berry-Robnik-Brody in the case
of dynamical localization) statistics is observed. The full spectrum
is a linear superposition of the spectrum of the chaotic eigenstates
and the spectrum of the regular eigenstates. The relative contributions
of the separated spectra in the Berry-Robnik(-Brody) statistics are
given by the fractional Liouville measure of the chaotic component
$\rho_{C}$ on the classical energy surface (i.e. the energy surface
of the equivalent classical system) and the measure of the regular
components $\rho_{R}=1-\rho_{C}$. Thus far we were involved with
determining the (fractional) measure of the chaotic component $\chi_{C}$
on the Poincaré surface of section (SOS). This is related to the fractional
Liouville measure on the energy surface via the following relation
\citep{Meyer1986} 
\begin{equation}
\rho_{C}=\frac{\chi_{C}}{\chi_{C}+\left(1-\chi_{C}\right)\xi},\label{eq:Rho}
\end{equation}
where $\xi=\frac{\left\langle t\right\rangle _{R}}{\left\langle t\right\rangle _{C}}$
is the ratio between the average return time to the SOS on the regular
components $\left\langle t\right\rangle _{R}$ and the average of
the same quantity on the chaotic component $\left\langle t\right\rangle _{C}$.
For the equivalent formula for $\rho_{R}=1-\rho_{C}$ we only need
to exchange the indices $R\leftrightarrow C$ and invert the ratio
$\xi\rightarrow1/\xi$ in Eq. \eqref{eq:Rho}. In billiards the SOS
is the billiard boundary and the ratio $\xi$ is independent of the
speed of the particle. The SOS return time is proportional to the
length of a link of a trajectory. The ratio $\xi$ can thus be determined
by averaging the length of a link over a number of collisions and
calculating the averages with regard to the initial conditions. The
measures of the chaotic component on the SOS as well as the associated
Liouville measures on the energy surface are given in table \ref{tab:Measure}
for various $\lambda$.

\begin{table}[h]
\begin{centering}
\begin{tabular}{|c|c|c|c|}
\hline 
$\lambda$ & $\chi_{C}$ & $\chi_{A}$ & $\rho_{C}$\tabularnewline
\hline 
\hline 
0.1175 & 0.382 & 0.397 & 0.452 \tabularnewline
\hline 
0.125 & 0.548 & 0.569 & 0.613\tabularnewline
\hline 
0.13 & 0.600 & 0.611 & 0.662\tabularnewline
\hline 
0.135 & 0.662 & 0.674 & 0.718\tabularnewline
\hline 
0.14 & 0.707 & 0.717 & 0.755\tabularnewline
\hline 
0.15 & 0.789 & 0.798 & 0.824\tabularnewline
\hline 
0.16 & 0.855 & 0.862 & 0.876\tabularnewline
\hline 
0.18 & 0.946 & 0.950 & 0.952\tabularnewline
\hline 
0.2 & 0.976 & 0.977 & 0.975\tabularnewline
\hline 
0.23 & 0.996 & 0.997 & 0.996\tabularnewline
\hline 
0.25 & 0.999 & 0.999 & 0.999\tabularnewline
\hline 
0.5 & 1 & 1 & 1\tabularnewline
\hline 
\end{tabular}
\par\end{centering}
\caption{\label{tab:Measure}Table of values for the fractional measure of
the largest chaotic component. $\chi_{C}$ is the value for the measure
on the SOS, obtained from fitting the occupancy distribution, $\chi_{A}$
is the proportion of filled cells at $T=10^{11}$ and $\rho_{C}$
is the measure on the energy surface calculated from $\chi_{C}$,
using Eq.\eqref{eq:Rho}. The cell grid size was $L=5000$.}
\end{table}

\section{\label{sec:Correlation}The correlation functions and momentum diffusion}

\begin{figure}[h]
\begin{centering}
\includegraphics[width=1\columnwidth]{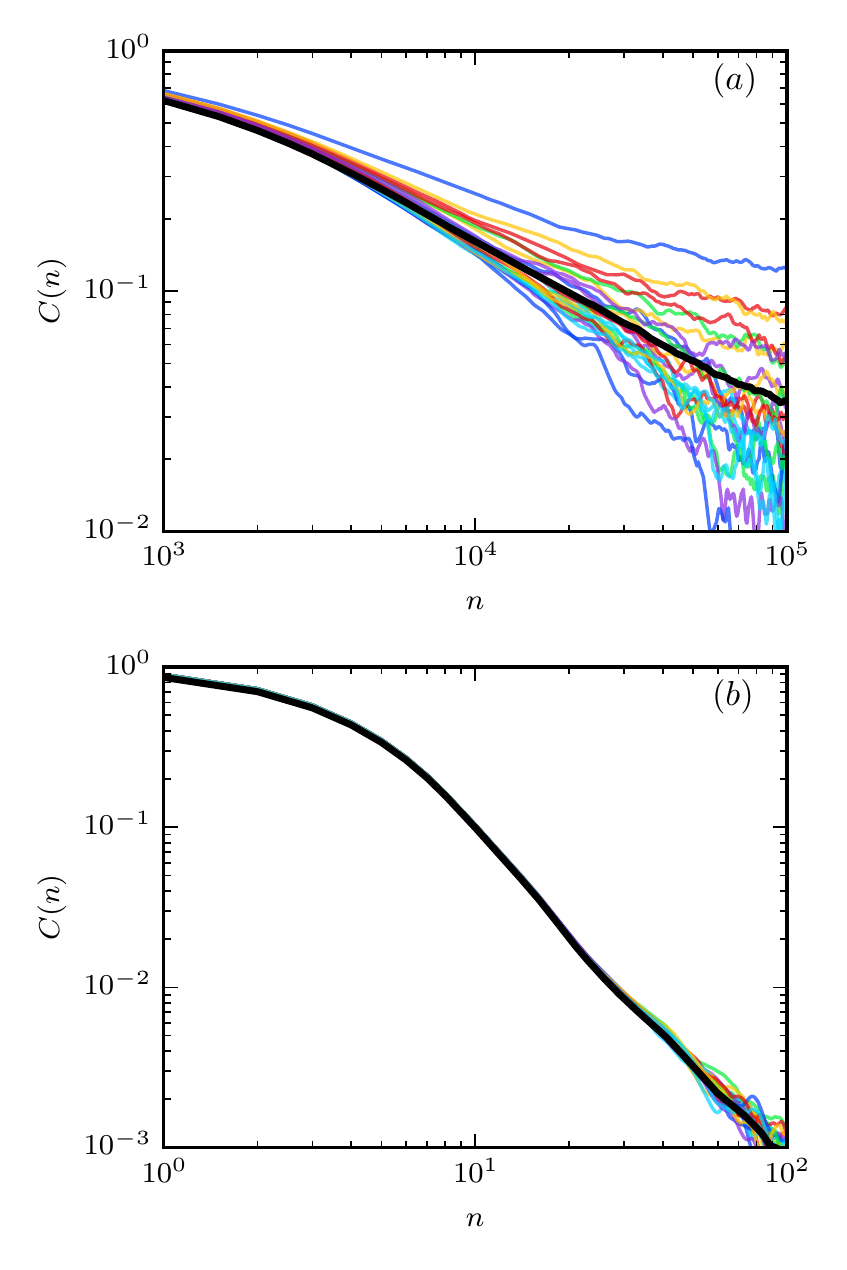}
\par\end{centering}
\caption{\label{fig:Correlations}The normalized momentum autocorrelation functions
in the $\lambda=0.15$ billiard (a) and the ergodic $\lambda=0.5$
billiard (b). The colored lines depict the autocorrelation functions
for 24 initial conditions selected randomly in the largest chaotic
component. The black line is the average over all 24 initial conditions.
The orbits were iterated $T=10^{8}$ times for the calculation of
the autocorrelation functions. }

\end{figure}

The phenomenon of stickiness has been shown to lead to non-exponential
or power law decay of correlations. As a means of studying the decay
of correlations in our family of billiards we chose the momentum autocorrelation
function defined as
\begin{equation}
C_{pp}\left(n\right)=\lim_{T\rightarrow\infty}\frac{1}{T}\sum_{t=0}^{T}p\left(x_{t}\right)p\left(x_{t+n}\right),
\end{equation}
where $p\left(x_{i}\right)$ denotes the momentum of the particle
after the $i$-th mapping of the orbit $x_{i}$ with initial condition
$x_{0}$. It is convenient to normalize the autocorrelation function
by its initial value $C\left(n\right)=\frac{C_{pp}\left(n\right)}{C_{pp}\left(0\right)}$.
The decay of the normalized momentum autocorrelation functions is
depicted in Fig. \ref{fig:Correlations}. Panel (a) shows $C\left(n\right)$
for 24 orbits in the $\lambda=0.15$ billiard. The initial conditions
were selected at random inside the largest chaotic component. Some
orbits start deep inside the chaotic sea while others start near the
various phase space structures like the KAM islands or near the Lazutkin
torus. We see that the decay of correlations differs significantly
from orbit to orbit. The average over all 24 initial conditions is
shown with the black line. The tail of this average obeys a power
law (linear in the log-log scale) $C\left(n\right)\sim n^{-\alpha}$
with $\alpha\approx0.68$. Panel (b) shows $C\left(n\right)$ for
24 orbits in the $\lambda=0.5$ billiard i.e. the ergodic case. The
decay of correlations here is very rapid with $C\left(n\right)$ reaching
the 0.1 value in about ten orbit iterations. The different initial
conditions produce practically the same results with the small differences
being attributed to numerical fluctuations. The tail of the average
also decays as a power with $\alpha=2$. This may be considered counter
intuitive as the system is ergodic and, in contrast to some other
prominent billiards like the stadium, has no families of marginally
unstable periodic orbits, that have been shown to produce such effects
\citep{VCG1983}. But it has also been shown that in the stadium billiard
the power law tails of the correlations are caused also by the orbits
only glancing the boundary and spending a significant number of iterations
in integrable motion on one of the stadium semicircles. These have
been specifically associated with a power law decay with $\alpha=2$.
While there is no complete analogy for this type of integrable motion
in the $\lambda=0.5$ billiard an orbit with a very large angle of
incidence may still spend a significant number of bounces glancing the
boundary with the angle changing only slightly and thus sticking to
the edge of the phase space until the inevitable flyaway, when it
reaches the cusp singularity at $z=-1$. It is important to note that
if we were to study the billiard in continuous time i.e. the billiard
flow instead of the billiard map, the quick succession of the bounces
would eliminate this type of stickiness. 

\begin{figure}[h]
\begin{centering}
\includegraphics[width=1\columnwidth]{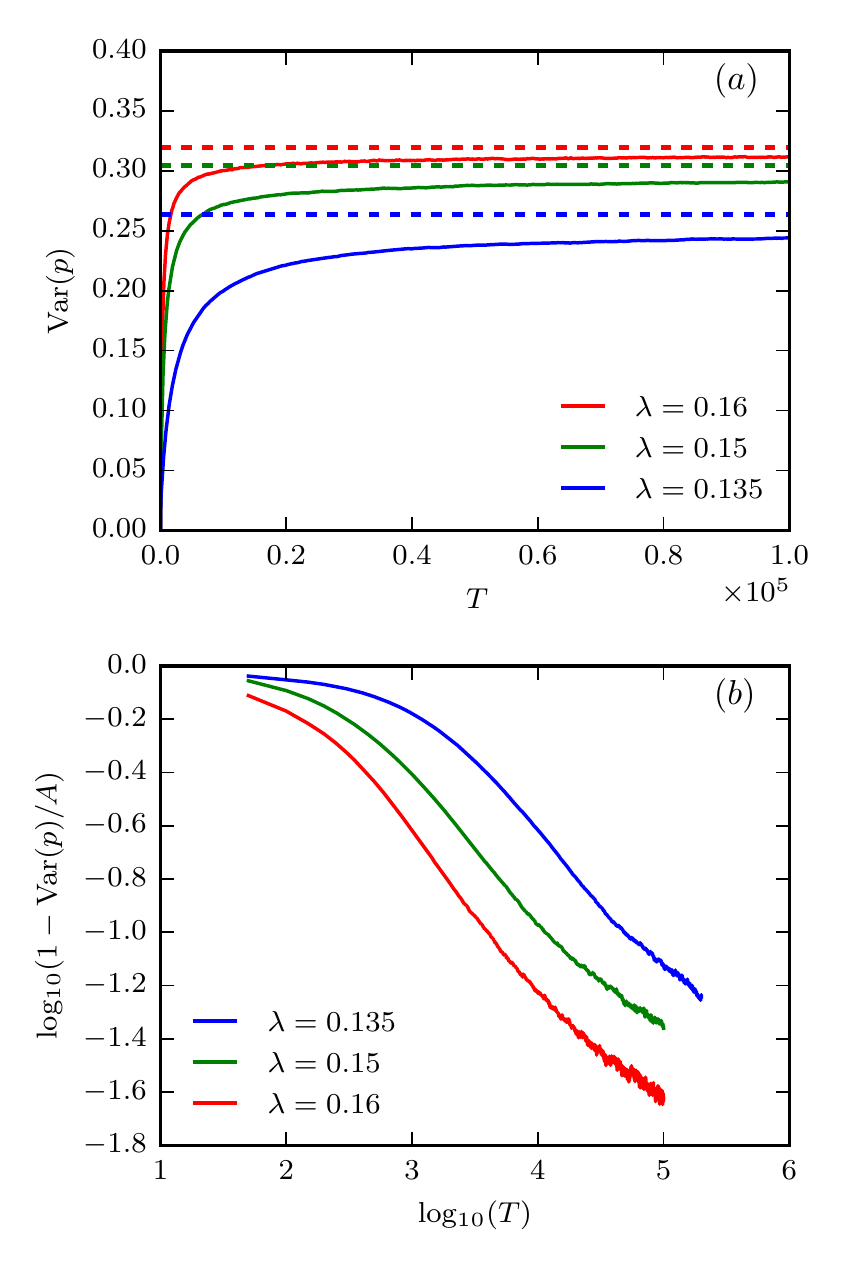}
\par\end{centering}
\caption{\label{fig:Diffusion}Momentum diffusion in the $\lambda=0.135$,
$\lambda=0.15$ and $\lambda=0.16$ billiards. (a) The variance
of the momenta of an ensemble of $10^{5}$ particles as a function
of the number of bounces. The particles all have an initial momentum
of $p=0$ and $s$ from the chaotic component. The expected asymptotic value of the variance, calculated from the phase portrait, is plotted with a dashed
line. (b) The approach to the asymptotic value in the log-log scale (the order of the curves from top to bottom is indicated by the legend in each panel).}

\end{figure}

Lastly we turn our attention to the diffusion in momentum space. In
a previous paper \citep{LozRob2018} we explored the momentum diffusion
of ensembles of particles in the stadium billiard and found that the
diffusion was well described by an inhomogeneous diffusion model with
a diffusion constant that is a quadratic function of the momentum.
We have shown that the variance of the momenta of an ensemble of particles
distributed along the billiard boundary and with initial momentum
$p_{0}$ approaches its asymptotic value exponentially. A similar
numerical experiment may be performed for the family of $\lambda$
billiards. Firstly, we note that momentum diffusion may only be present
on the chaotic component and not on the regular ones. We thus selected
an ensemble of $10^{5}$ particles with initial conditions $p=0$
(the initial velocity of the particles is perpendicular to the boundary)
and $s$ in the chaotic sea. The particles then bounce inside the
billiard and the variance of their momenta $\mathrm{Var}(p)$ is recorded
with each bounce. Eventually, the particles should distribute themselves
uniformly across the chaotic component and the asymptotic value of
the variance $A$ should be reached. In contrast to the stadium billiard
the value of $A$ can not be easily obtained analytically due to the
complex structure of the phase space (the exception is $\lambda=0.5$).
The value of $A$ was determined numerically from the phase portraits
at $L=5000$ and should be accurate within one percent. In Fig. \ref{fig:Diffusion}
the results for $\lambda=0.135,$ 0.15 and 0.16, with $p_{0}=0$ are
shown. Panel (a) shows the variance of the momentum as a function
of the number of bounces. The results for all three $\lambda$ are
qualitatively similar. An initial rapid increase of the variance is
followed by a slow approach to the asymptotic value. This is quite
different to the case of the stadium where the variance was globally
well described by an exponential decay to the asymptotic value. This
slow decay to the asymptotic value is probably a consequence of the
labyrinthine complexity of the phase space and the trapping of orbits
by sticky objects. The approach to the asymptotic value in the log-log
scale is portrayed on panel (b). The approach to the asymptotic value
seems to be neither exponential nor a power law.

\section{\label{sec:Discussion}Discussion and conclusions}

We have shown that in generic (mixed-type) Hamiltonian systems the
complexity of the phase space and the intermittency of the chaotic
dynamics pose a serious challenge when determining the size and structure
of the chaotic component. When the phase space is divided into cells
the cell visitations by a chaotic orbit are not completely uncorrelated
even for very small cells. The orbit may be trapped by sticky objects
or cantori for significant periods of time. The standard deviation
of the cell recurrence time may be a good observable for quantifying
stickiness on the global level. The standard deviation of the cell
recurrence times decays roughly exponentially with the distance to
the sticky object. The vast majority of cells in the chaotic component
experience completely uncorrelated visits, as evidenced also by the
recurrence time distributions. In the long run the intermittent behavior
of the chaotic orbit averages out and the cell visitation statistics
concur with the random model. From there one can extract the measure
of the chaotic component, either from the distribution of the cell
visitation numbers or the average recurrence times, to a high degree
of accuracy. 

This method is an improvement over simply counting the number of chaotic
cells and is less dependent on cell size. However, we must stress
that the number of orbit iterations needed for the desired accuracy
may still be very large, in our case $10^{11}$. A nice feature of
the method is also that the only prior knowledge of the system we
need is one initial condition in the chaotic component we wish to
measure. The method should be applicable to a wide range of dynamical
systems and is a priori not limited to two dimensions, although the
scaling of the cell volume might make its use prohibitive in higher
dimensions. One of the ways in which this problem might be addressed
is by computing the cell visits of several chaotic orbits in parallel
and combining the results.

As has been observed in previous examples, stickiness leads to power
law decay of correlations also in our family of billiards. The momentum
diffusion also behaves differently than in the stadium billiard where
the approach of the variance of the momenta to
the asymptotic value is exponential. The classical transport time
(or diffusion time) is an important parameter in the study of dynamical
localization in quantum chaos. In the stadium it was possible to define
the classical transport time in a clear way, by using this exponential
law. It is not possible to define the classical transport time for
the family of billiards used in this paper in the same way and this
remains an open question. Our current working definition involves
measuring the time it takes for the variance to reach an arbitrary
proportion of the asymptotic value and further work is in progress.

\section*{Acknowledgments}

The authors acknowledge the financial support from the Slovenian Research
Agency (research core funding No. P1-0306). We would like to thank
Dr. B. Batisti\'c for stimulating discussions and providing the use
of his excellent numerical library \citep{Benokit}. 

\bibliographystyle{apsrev4-1}
\nocite{*}
\providecommand{\noopsort}[1]{}\providecommand{\singleletter}[1]{#1}%

\end{document}